\documentclass[amsmath,amssymb,aps]{revtex4}
\usepackage{graphicx}% Include figure files
\usepackage{dcolumn}% Align table columns on decimal point
\usepackage{bm}% bold math

\begin{document}

%\preprint{}
%APS/123-QED

\title{The Principle of Strain Reconstruction Tomography: \\
Determination of Quench Strain Distribution from Diffraction Measurements}

\author{Alexander M. Korsunsky}
\email{alexander.korsunsky@eng.ox.ac.uk}
\affiliation{Department of Engineering Science, University of Oxford\\ 
Parks Road, Oxford OX1 3PJ, England}

\author{Willem J.J. Vorster}
\affiliation{Department of Engineering Science, University of Oxford\\ 
Parks Road, Oxford OX1 3PJ, England}

\author{Shu Yan Zhang}
\affiliation{Department of Engineering Science, University of Oxford\\ 
Parks Road, Oxford OX1 3PJ, England}

\author{Daniele Dini}
\affiliation{Department of Engineering Science, University of Oxford\\ 
Parks Road, Oxford OX1 3PJ, England}

\author{David Latham}
\affiliation{Department of Engineering Science, University of Oxford\\ 
Parks Road, Oxford OX1 3PJ, England}

\author{Mina Golshan}
\affiliation{Synchrotron Radiation Source, Daresbury Laboratory
Keckwick Lane, Warrington WA4 4AD, England}

\author{Jian Liu}
\affiliation{Department of Chemistry, University of Durham
South Road, Durham DH1 3LE, England}

%\author{Ioannis Kyriakoglou}
%\affiliation{Rolls-Royce plc, P.O. Box 31, Derby DE24 8BJ, England}

%\author{Michael J. Walsh}
%\affiliation{Rolls-Royce plc, P.O. Box 31, Derby DE24 8BJ, England}

\date{\today}% It is always \today, today,

%\pacs{62.20.Fe, 62.20.Mk, 62.20.Qp, 62.20.Hg}% PACS, the Physics and Astronomy
\begin{abstract}

Evaluation of residual elastic strain within the bulk of engineering components or natural objects is a challenging task, since in general it requires mapping a six-component tensor quantity in three dimensions. A further challenge concerns the interpretation of finite resolution data in a way that is commensurate and non-contradictory with respect to continuum deformation models. A practical solution for this problem, if it is ever to be found, must include efficient measurement interpretation and data reduction techniques. In the present note we describe the principle of strain tomography by high energy X-ray diffraction, i.e. of reconstruction of the higher dimensional distribution of strain within an object from reduced dimension measurements; and illustrate the application of this principle to a simple case of reconstruction of an axisymmetric residual strain state induced in a cylindrical sample by quenching. The underlying principle of the analysis method presented in this paper can be readily generalised to more complex situations.

\end{abstract}

\maketitle

\section{\label{sec:intro}Introduction}

The diffraction of neutrons and synchrotron X-rays provides a unique non-destructive probe for obtaining information on strains deep in the bulk of engineering components and structures. It has become a mature tool for the determination of residual strain states in small coupons, and developments are under way to establish the facilities for performing high resolution measurements directly on larger engineering components.

There are two principal methods for extracting residual elastic strain information within objects with the help of diffraction, namely, the angle dispersive (monochromatic beam) technique, and the energy-dispersive (white-beam) technique. 

In the monochromatic mode the beam is first passed through an optical element that transmits only photons with the energy or wavelength lying within a narrow bandwidth. The optical element in question (the monochromator) is usually a highly perfect crystal placed in the beam in reflection orientation (Bragg mode) or transmission orientation (Laue mode). The band pass depends on the quality of the crystal, and can be as narrow as $\rm10^{-4}$ or better. In many practical situations this accuracy may exceed requirements, and broader band pass is in fact beneficial, since it increases flux on the sample, the number of crystals within a polycrystalline sample that contribute to the pattern, etc. 

Diffraction patterns are collected by employing a detector scanning the scattering angle 2q, or a position sensitive detector capable of registering total photon flux simultaneously at several positions along a line or over a two-dimensional surface. This mode allows accurate determination of diffraction peak intensity, shape and position. However, it usually requires significantly longer counting times in comparison with the white beam mode in order to collect the data from comparable sections of the diffraction pattern, primarily due to the reduction of flux by monochromation, but also due to the necessity of scanning the detector.

Energy dispersive setup allows multiple diffraction peaks to be collected simultaneously, thus achieving particularly efficient counting statistics at energies above 30 keV \cite{JSR}. The accuracy of determination of individual peak position and shape resolution in the white-beam mode is usually related to the resolution of the energy-dispersive detector, but can in fact be several orders of magnitude better \cite{liu}. The accuracy of interpretation in terms of lattice parameters and hence strain can be significantly improved by using multiple peak analysis or whole pattern fitting \cite{liu}.

A particular feature of high energy X-ray diffraction is that the radiation is primarily scattered forward, i.e. in directions close to that of the incident beam. Therefore small diffraction angles have to be used, usually $2\theta<15^\circ$. Two difficulties immediately follow. 

Firstly, it is difficult to measure strains in directions close to that of the incident beam. This is due to the fact that the scattering vector is always the bisector of the incident and diffracted beams. Hence for high energy X-rays the strain measurement directions form a shallow cone. For a scattering angle of $2\theta$  this cone has the angle of $(180^\circ-2\theta)/2=90^\circ-\theta$. In practice this means that strain directions accessible for the high energy X-ray diffraction techniques are close to being normal to the incident beam. This situation is in stark contrast with that encountered in laboratory X-ray diffraction where near backscattering geometry is used, and measured strains are in directions close to being parallel with the incident beam.

\begin{figure}
\centerline{ \includegraphics[width=13.cm]{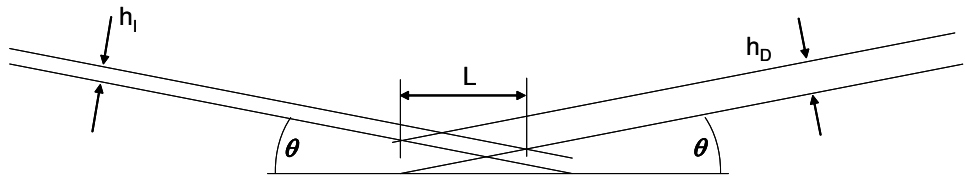} }
\caption{Geometry of the incident and diffracted beams and the gauge volume 'lozenge'.}
\label{fig:one}
\end{figure}

Secondly, because of the small scattering angle the gauge volume becomes a strongly elongated lozenge (Fig.\ref{fig:one}). The total length of the scattering gauge volume is
\begin{equation}
L=(h_I+h_D)\cos\theta/\sin 2\theta=(h_I+h_D)/(2\sin\theta),
\label{eq:gauge}
\end{equation}
where $L$ is the length of the gauge volume, $h_I$ denotes the height of the incident beam, and $h_D$ denotes the height of the diffracted beam. If these two heights a both equal to $h$, then equation (\ref{eq:gauge}) simplifies to 
\begin{equation}
L/h=1/sin\theta
\label{eq:ga}
\end{equation}
For a scattering angle $2\theta$ of $10.7^\circ$ the ratio $L/h$ is close to 10.7. As a consequence the resolution in the direction of the incident and scattered beam is found to be degraded about ten-fold compared to the resolution in the perpendicular plane.

The motivation for the present paper lies in the desire to overcome these limitations by combining the information available from multiple measurements. In many respects the situation is similar to that encountered in reconstruction tomography, when a three dimensional structure is de-convoluted, or, rather, re-constituted from 2D projections. There is an important distinction between the approach proposed here and the algorithms that make use of back-projection, but are ignorant of the properties of the object. Here instead we propose to use a parametric model of the object - in the present case, the unknown strain distribution - and to carry out a minimisation procedure on the mismatch between the measurements and their simulation. In the present paper we deliberately choose a very simple distribution of residual elastic strain, since this allows us to introduce the mathematical without obfuscating it with too many details. However, the same fundamental approach can be applied to more complex situations, and this work is under way.

The use of diffraction strain analysis for residual stress characterisation holds out the prospect of furnishing an important validation tool for tuning up manufacturing processes so as to reduce distortion and residual stress, and to increase the durability of components. In recent years powerful novel modelling approaches have been put forward for the analysis of residual elastic strain distributions and the corresponding residual stresses, most notably using the eigenstrain modelling approach \cite{JSA}.

Heat treatment procedures play a significant role in controlling the metallurgical processes within components that deliver the desired alloy properties in terms of the combination of strength, hardness, toughness, ductility, etc. Quenching is often used in order to provide the rate of cooling required in order to form precipitation-hardened microstructures, to create metastable phases with particular properties, such as martensite, etc. However, invariably quenching leads to the creation of high thermal gradients that result in differential contraction or expansion of neighbouring material volumes. This, accompanied by the strong dependence of yield stress on temperature, usually observed in metallic alloys, leads to ready plastic deformation in the hotter parts of the specimen or component subjected to quenching treatment, i.e. within the bulk of material and away from the surface. As a result of the plastic strain distribution 'frozen' into the material, a residual elastic strain (r.e.s.) and the corresponding residual stress distribution arise. 

These can be beneficial or detrimental to the service properties of the component, depending on the loading it experiences in service. For example, the usual result of the quenching procedure is to create a state of residual compression near the surface, which helps to inhibit the initiation and propagation of fatigue cracks. This phenomenon is used in quench and induction hardening processes. 

For the present study we have chosen the residual elastic strain state in a quenched cylinder of nickel superalloy IN718. The residual strain state is known to be axisymmetric, and in fact is known to follow a parabola as a function of radial distance from the axis. These elements of knowledge help formulate a simple model for the unknown strain distribution, and simplify the reconstruction procedure considerably. 

\section{\label{sec:exp}Experimental}

The sample used for the experiment was made from IN718 creep resistant nickel superalloy used in the manufacture of aeroengine components, such as combustion casings and liners, as well as disks and blades. The composition of the alloy in weight percent is approximately given by 53\% Ni, 19\% Fe, 18\% Cr, 5\% Nb, and small amounts of additional alloying elements Ti, Mo, Co, and Al. Apart from the matrix phase, referred to as $\gamma$, the microstructure of the alloy may show additional precipitate phases, referred to as $\gamma'$, $\gamma''$, and $\delta$.
 
The primary strengthening phase, $\gamma''$, has the composition $\rm Ni_3 Nb$ and a body-centred tetragonal structure, and forms semi-coherently as disc-shaped platelets within the $\gamma$ matrix. It is highly stable at $600^\circ$C, but above this temperature it decomposes to form the $\gamma'$ $\rm Ni_3 Al$ phase (between $650^\circ$C and $850^\circ$C), and $\delta$, having the same composition as $\gamma''$ (between $750^\circ$C and $1000^\circ$C). At large volume fractions and when formed continuously along grain boundaries, the $\delta$ is thought to be detrimental to both strength and toughness \cite{Brooks}. The $\delta$ phase that forms is more stable than the $\gamma''$ phase, and has an orthorhombic structure \cite{Guest}.

The sample was made into the form of a cylinder of a=6.5mm in diameter and was subjected to high temperature solution treatment for two hours at $920^\circ$C, followed by quenching axially into cold water. This treatment results in the creation of an axially symmetric residual elastic strain profile which varies roughly according to the parabolic law with the distance from the sample axis, with the surface regions of the sample experiencing compressive axial residual elastic strain, and its core tensile axial residual elastic strain. As follows from the above discussion of IN718 metallurgy, we may expect to find at least three phases present in the material following this heat treatment, the matrix $\gamma$ phase together with $\gamma''$ and $\delta$ phases, and possibly some traces of $\gamma'$.

The sample was mounted on the Euler cradle on Station 16.3 at the synchrotron radiation source at Daresbury Laboratory. Apart from providing the usual rotational degrees of freedom $(\omega, \chi, \phi)$ the cradle is equipped with two translators allowing movement in the plane of the sample support platform. 

White beam energy dispersive set up was used to collect the data. The station insertion device, wavelength shifter, produces a smooth white beam spectrum with useable photons having energies up to 100keV. Only photons with energies exceeding 60keV were useful in the present experiment. The diffraction spectrum was recorded using a Li-drifted Ge detector (Canberra) that was cooled with liquid nitrogen. A multi-channel analyser was used to bin the pulses from the amplifier.

Detector calibration was carried out using the procedure described by Liu et al \cite{liu}, firstly using a radioactive source, then using the diffraction pattern collected from the standard NIST silicon powder sample.

\begin{figure}
\centerline{ \includegraphics[width=16.cm]{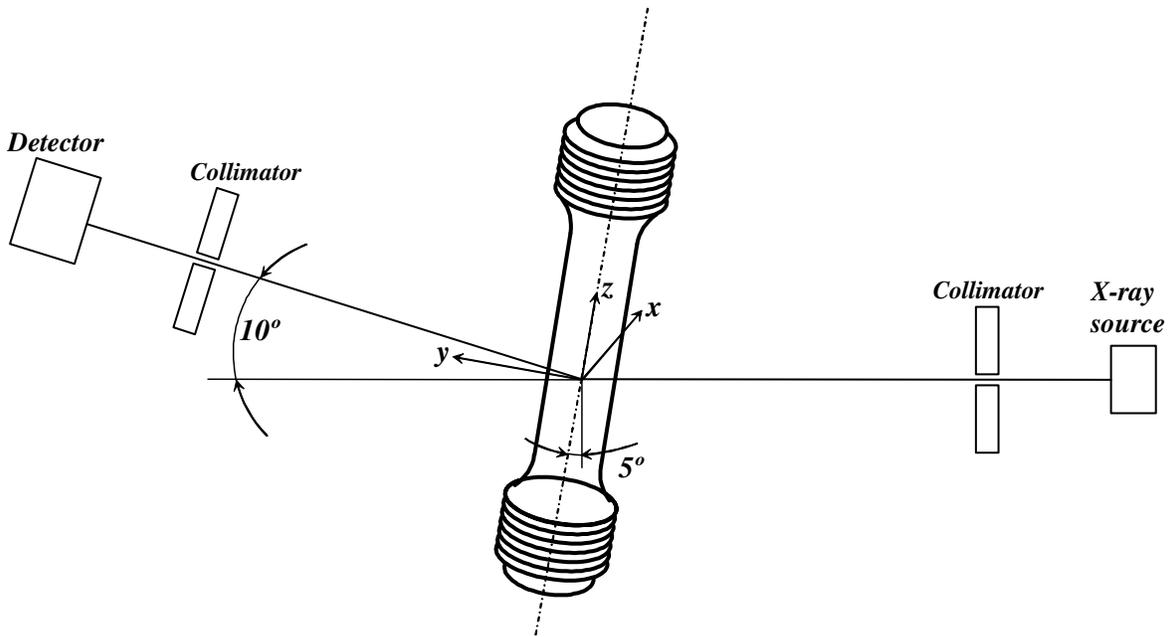} }
\caption{Schematic diagram of the experimental setup.}
\label{fig:two}
\end{figure}

The sample was mounted as shown in Fig.\ref{fig:two}. The translation $x$-axis perpendicular to the incident beam direction was located on the stage attached to the Euler cradle. The stage was tilted around the axis perpendicular to the incident beam by the angle of $5^\circ$, i.e. half the scattering angle of $10^\circ$. Slit apertures of 1.2mm for the incident beam and 0.1mm for the diffracted beam were used in the experiment, producing the gauge volume of the type shown in Fig.\ref{fig:one}. The symmetry axis of the cylindrical sample was designated as $z$-axis, and the $y$-axis was normal to both $x$ and $z$ directions and formed a right handed coordinate system.

\section{\label{sec:diff}Diffraction data interpretation}

In order to determine the average value of the lattice parameter within each gauge volume, Rietveld refinement is carried out on the diffraction patterns using GSAS. 

Our preferred data treatment procedure has been described by J. Liu et al. \cite{liu} and involves careful stepwise calibration of the detector characteristic, followed by progressive improvement of the fit quality. Detector calibration begins with the use of an Am209 radioactive source that provides photon flux at three distinct energies, although two of these energies are relatively close to each other, approximately at 15, 20 and 60keV. This is insufficient to determine the non-linear detector characteristic, but allows approximate linear calibration. At the next stage a diffraction pattern is used that was collected from a powder of a standard material (e.g. NIST Si) with precisely known lattice parameter. Starting with the previously known linear detector calibration, quadratic characteristic of the detector is sought by requiring the best fit to the diffraction pattern. It is important to note, in passing, that if the diffraction angle is not known precisely, it can be determined at this stage together with the channel-to-energy conversion.

Pattern refinement began with the primary fcc gamma phase in the nickel superalloy IN718. This is known to co-exist in coherent form with the gamma-prime phase. It was found, however, that this did not provide an adequate fit to the experimental diffraction pattern, particularly with the peaks at 71 and 73 keV remaining unaccounted for. 

\begin{figure}
\centerline{ \includegraphics[width=16.cm]{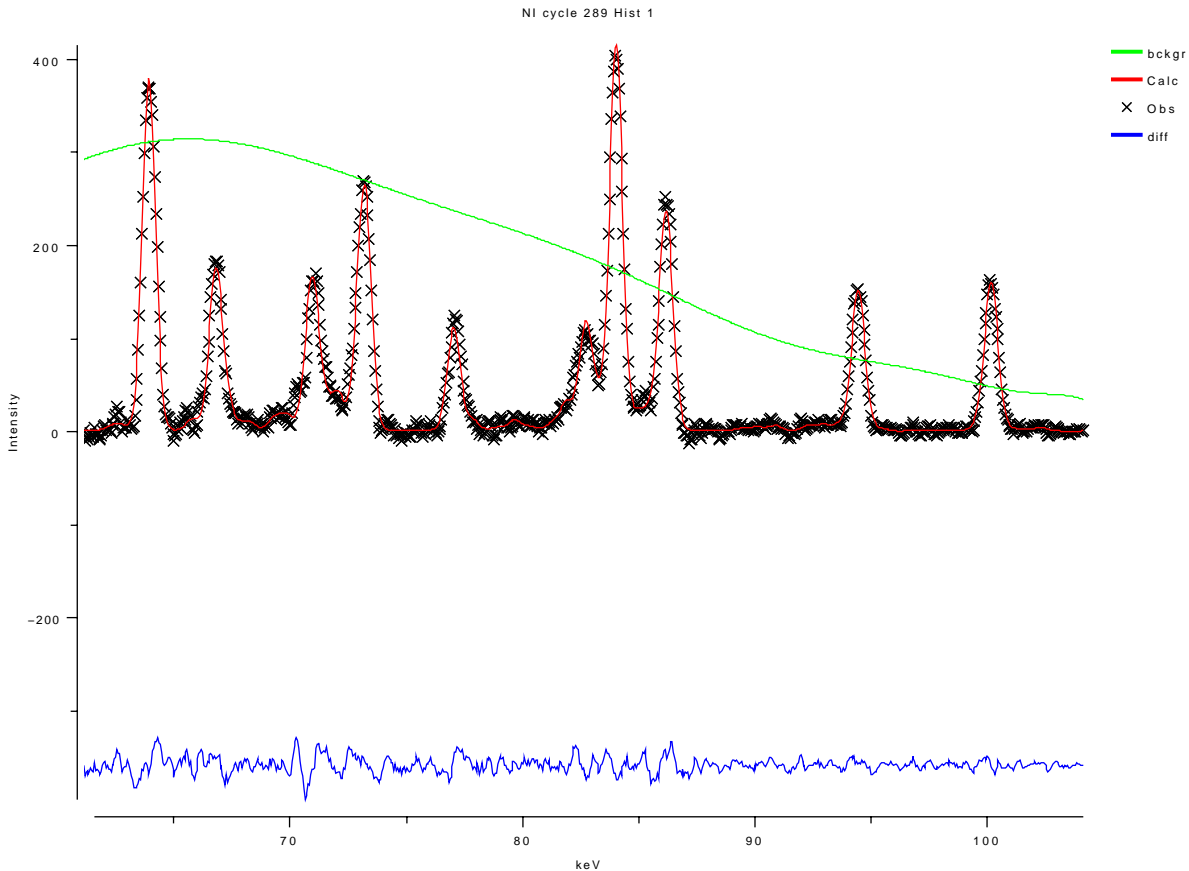} }
\caption{Illustration of the GSAS fit to the diffraction pattern. Three phases used in the fit (gamma, gamma prime and orthorhombic) allow excellent description of the data, as shown by the difference plot (bottom curve). The diffraction pattern (cross markers and continuous fit curve) is shown with the background (smooth curve) subtracted. The background was quite high in this experiment, since measurements were made at the higher limit of the energy range accessible on station 16.3 at the SRS.}
\label{fig:three}
\end{figure}

Fig.\ref{fig:three} provides an illustration of the GSAS fit to the diffraction pattern. The use of two phases in the fit ($\gamma$ and $\delta$) allows excellent description of the data, as shown by the difference plot (bottom curve). The diffraction pattern (cross markers and continuous fit curve) is shown with the background (smooth curve) subtracted. The background was quite high in this experiment, since measurements were made at the higher limit of the energy range accessible on station 16.3 at the SRS.

The a parameter of the majority matrix $\gamma$ phase was used in the calculation of the lattice residual elastic strain according to the usual formula, 
\begin{equation}
\epsilon=\frac{a}{a_0}-1,
\label{eq:strain}
\end{equation}
where $a_0$  denotes the reference 'strain-free' value of the lattice parameter. We do not presuppose this value, but instead carry the above formula in 'dynamic' form through our calculations below, i.e. retain   as a parameter within the computation, thus being able to adjust it at the end to satisfy the conditions of equilibrium.

\section{\label{sec:tomo}Strain reconstruction tomography}

The approach adopted in this study rests on the postulate that it is necessary to make informed assumptions about the nature of the object (or distribution) that is being determined in order to formulate and regularise the problem. Interpretation of any and every measurement result incorporates a model of the object being studied. To give a very general example, when a ruler is pressed against the sample to determine its length, the implication is that the sample and ruler surfaces are collocated all along the measured length. If that is so, then the reading from the ruler is correct; otherwise, however, the measurement will be a lower estimate.

As an example of more direct relevance to the present subject, consider the situation that arises when the residual stress within the sample surface is 'measured' using the $\sin^2\psi$ method, a whole array of assumptions is being made, namely (i) that the material is uniform, isotropic and continuous, (ii) that strain values measured at different angles of tilt, $\psi$, are obtained from the same group of grains within the same gauge volume; (iii) that the component of stress normal to the sample surface vanishes within the gauge volume; etc. All of the above assumptions are in fact approximations, or, in worst cases, entirely wrong.

In developing the mathematical technique presented below we propose to go one step further and adopt the approach that it is preferable to develop measurement interpretation procedures that explicitly incorporate a model of the measured object. This approach would seem particularly appropriate in cases when each individual measurement (each data point) in fact represents a convolution of the distribution being investigated, e.g. the average value of residual elastic strain being measured over the diffraction gauge volume.

In the case in hand, the unknown distribution of axial residual elastic strain (r.e.s.) is known to possess a strong symmetry property due to the nature of the sample geometry. Due to its axial symmetry, r.e.s. can be modelled using a function of radial position only. For ease of analysis and without significant loss of generality, we represent the unknown strain distribution as a function of radial position within the sample by the following truncated power series:
\begin{equation}
e(r)=\sum_{i=0}^n c_i e_i(r)=\sum_{i=0}^n c_i r^i, \quad r\leq a,
\label{eq:trunc}
\end{equation}

Here $a$ denotes the radius of the cylindrical specimen, $n+1$ is the total number of terms, and $c_i$ denotes the unknown coefficients, yet to be determined, within the series of radial power functions, $e_i(r)=r^i$. The cylindrical polar system of coordinates $(r,\theta,z)$ is used, for which the $z$-axis coincides with the axis of symmetry of the quenched cylinder. For the purposes of subsequent discussion we shall also write the same expression in the Cartesian coordinate system $(x,y,z)$ which shares the $z$ axis with the polar system, so that
\begin{equation}
e(x,y)=\sum_{i=0}^n c_i \left( \sqrt{x^2+y^2} \right)^i.
\label{eq:truncart}
\end{equation}

We now proceed to simulating the result of diffraction measurements. In the experiments described in the present study the incident beam height amounted to 1.2mm and the diffracted beam height to 0.1mm. From equation (\ref{eq:gauge}) it follows that, for scattering angle of $2\theta =10^\circ$, the length of the gauge volume therefore amounted to 7.5mm and spanned the full depth of the sample at all positions. It is also possible to note, from considering Fig.\ref{fig:one} in detail that the length of the gauge volume having uniform height is given by
\begin{equation}
L=h_I/(2\sin\theta),
\end{equation}
and is equal to 6.9mm in the case considered. Thus, the constant height part of the scattering volume spanned the full depth of the section of the sample made by the plane $z=\rm const$, Fig.\ref{fig:two}.

\begin{figure}
\centerline{ \includegraphics[width=16.cm]{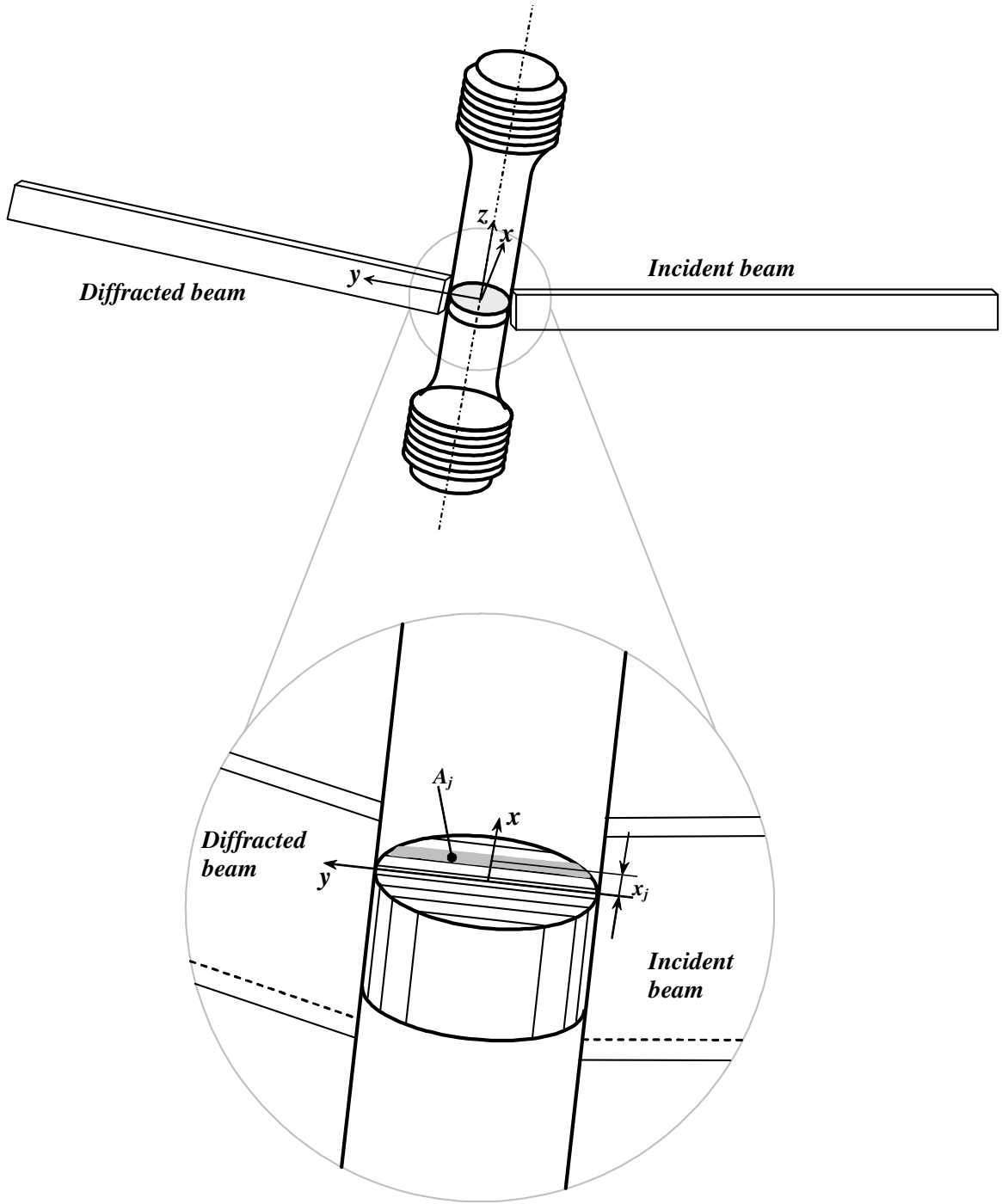} }
\caption{Illustration of the gauge volume positions with respect to the specimen.}
\label{fig:four}
\end{figure}

The next significant assumption made in our interpretation was that the strain value obtained from the diffraction profile at each point in the $x$-scan (Fig.\ref{fig:four}) is given by the average value of strain within the sampling volume,
\begin{equation}
\overline{e}(x_j)=\frac{\iint_{A_i} e(r){\rm d}A}{\iint_{A_i} {\rm d}A}.
\label{eq:ave}
\end{equation}

Here residual elastic strain in the sample, $e$, is assumed to be a function of radial position $r$ only, ${\rm d}A$ refers to the element of cross-sectional area of the sample. Area  corresponds to the 'footprint' of the scattering gauge volume within the sample centred, in the $x$ direction, at position $x_j$  within the sample,
\begin{equation}
A_j: \quad x_j-\Delta x/2 < z < x_j+\Delta x/2, \quad -\sqrt{a^2-x^2} < y < \sqrt{a^2-x^2}.
\label{eq:aj}
\end{equation}

Here $\Delta x$ refers to the beam width in the horizontal ($x$) direction that was set to 1mm in the experiment.

Using  $r=\sqrt{x^2+y^2}$, together with the definition of area  $A_j$  of equation (\ref{eq:aj}), allows equation (\ref{eq:ave}) to be used as a definition of the following matrix $\overline{e}_{ij}$:
\begin{equation}
\overline{e}_{ij}=\frac{\iint_{A_i} e_i(\sqrt{x^2+y^2}){\rm d}x {\rm d}y}{\iint_{A_i} {\rm d}x {\rm d}y}.
\label{eq:eij}
\end{equation}

Due to the problem's linearity, the predicted measured strains at points $x_j$ are given by
\begin{equation}
\overline{e}(x_j)=\sum_{i=0}^n c_i \overline{e}_{ij}.
\label{eq:ebar}
\end{equation}

We now form a measure of the goodness of fit provided by the model as the functional
\begin{equation}
G=\sum_{j=1}^K(\overline{e}(x_j)-\epsilon(x_j))^2=\sum_{j=1}^K \left( \sum_{i=0}^n c_i \overline{e}_{ij}-\epsilon(x_j) \right)^2,
\label{eq:G}
\end{equation}
where $\epsilon(x_j)$ denotes the strain measured at position $x_j$ by diffraction.

The search for the best choice of model can now be accomplished by minimising $G$ with respect to the unknown coefficients, $c_i$, i.e. by solving
\begin{equation}
{\rm grad}_{c_i} G=(\partial G/\partial c_i)=0, \quad i=1,...,n. 
\label{eq:grad}
\end{equation}

Due to the positive definiteness of the quadratic form (\ref{eq:G}), the system of linear equations (\ref{eq:grad}) always has a unique solution that corresponds to a minimum of $G$. 

The partial derivative of $G$ with respect to the coefficient $c_i$  can be written explicitly as
\begin{equation}
\partial G/\partial c_i = 2 \sum_{j=1}^K \overline{e}_{ij} \left( \sum_{k=0}^n c_k \overline{e}_{ij}-\epsilon(x_j) \right)
= 2\left( \sum_{k=0}^n c_k \sum_{j=1}^K \overline{e}_{ij} \overline{e}_{kj} - \sum_{j=1}^K \overline{e}_{ij}\epsilon(x_j) \right) = 0.
\label{eq:dgdc}
\end{equation}

We introduce the following matrix and vector notation
\begin{equation}
{\bf E} = \{ \overline{e}_{ij} \}, \quad {\bf\epsilon}=\{\epsilon(x_j)\}, \quad {\bf c}=\{ c_i\}.
\label{eq:matrix}
\end{equation}
Noting that in the context of equation (\ref{eq:dgdc}) notation $\overline{e}_{kj}$ corresponds to the transpose of the matrix $\bf E$, the entities appearing in equation (\ref{eq:matrix}) can be written in matrix form as:
\begin{equation}
{\bf A} = \sum_{j=1}^K \overline{e}_{ij} \overline{e}_{kj} = {\bf E E}^T, \quad 
{\bf b}=\sum_{j=1}^K \overline{e}_{ij}\epsilon(x_j) = {\bf E y}.
\label{eq:ab}
\end{equation}
Hence equation (\ref{eq:dgdc}) assumes the form
\begin{equation}
\triangle_{\bf c} G=2({\bf A c}-{\bf b})=0.
\label{eq:tri}
\end{equation}
The solution of the inverse problem has thus been reduced to the solution of the linear system
\begin{equation}
\bf A b = c
\label{eq:linsys}
\end{equation}
for the unknown vector of coefficients ${\bf c}=\{c_i\}$.

The interpretation procedure described above is most conveniently implemented in {\tt Mathematica} %$^\registered$ 
since it allows expressions such as equation (\ref{eq:ave}) involving integration or averaging commands to be coded in the form of functions, and vectors and matrices to be generated by the application of Listable functions or using the Table command.

We now turn to the selection of the particular strain distribution model, following the general form given in equation (\ref{eq:trunc}). We immediately note that the linear term must always be absent from the expression in equation (\ref{eq:trunc}) to avoid the appearance of physically unacceptable strain gradient discontinuity at $r=0$ on the cylinder axis. Due to the simplicity of the problem, the appropriate form of function in equation (\ref{eq:trunc}) describing the radial strain variation should be chosen in the form
\begin{equation}
e(r) = c_0-c_2 r^2, \quad r \leq a.
\label{eq:parab}
\end{equation}
In fact, tests have been performed with larger number of terms (up to 10), but as expected they did not produce an improvement in the quality of approximation.

In order to find the reference lattice spacing $a_0$ referred to in equation (\ref{eq:strain}), it is most desirable to apply the axial ($z$) stress balance condition. The values of radial strain were not available from the measurements, although a twin detector experimental setup has been proposed to achieve this measurement simultaneously with the axial strain \cite{epsrc}. In the present case, however, quench simulations demonstrate that the radial and hoop strains are much lower than the axial components, amounting only to about 10\% of the latter, with one component being tensile and the other compressive. As a consequence of this observation, the Poisson effect of transverse strains on the axial stress is not expected to exceed about 3\%. Consequently, axial strain balance can be applied instead of stress balance, in the form
\begin{equation}
B(a_0)=\int_0^a e(r) 2\pi r {\rm d} r = \pi (c_0 a^4 - c_2 a^4/2) = 0.
\label{eq:bal}
\end{equation}
It is possible to consider this equation as a fixed relationship between parameters $c_0$ and $c_2$. However, we use this expression as an implicit equation for the unknown unstrained lattice spacing $a_0$, thus enforcing strain balance.

\begin{figure}
\centerline{ \includegraphics[width=16.cm]{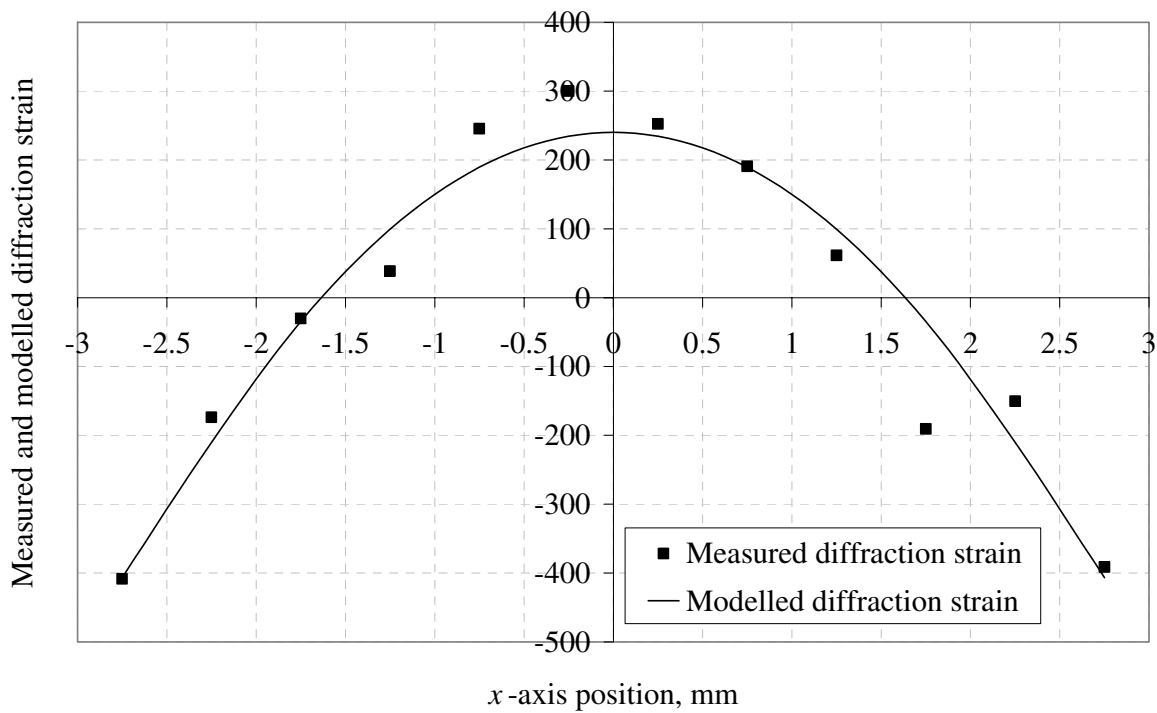} }
\caption{Comparison between the measured diffraction strain (markers) and the modelled diffraction strain (continuous curve), as a function of position within the x-scan centred on the sample axis of symmetry.}
\label{fig:five}
\end{figure}

\section{\label{sec:results} Results and discussion}

The results of the tomographic strain reconstruction are illustrated in Fig.\ref{fig:five} by way of comparison between the measured diffraction strain and the modelled strain, plotted as a function of position within the $x$-scan centred on the cylindrical axis of the sample. The quality of the agreement is clearly good, if a small amount of experimental scatter is taken into account. In fact, it may be stated that the agreement between the model and the measurements is in fact the best that can be delivered by a strain function obeying a quadratic variation profile. 

\begin{figure}
\centerline{ \includegraphics[width=16.cm]{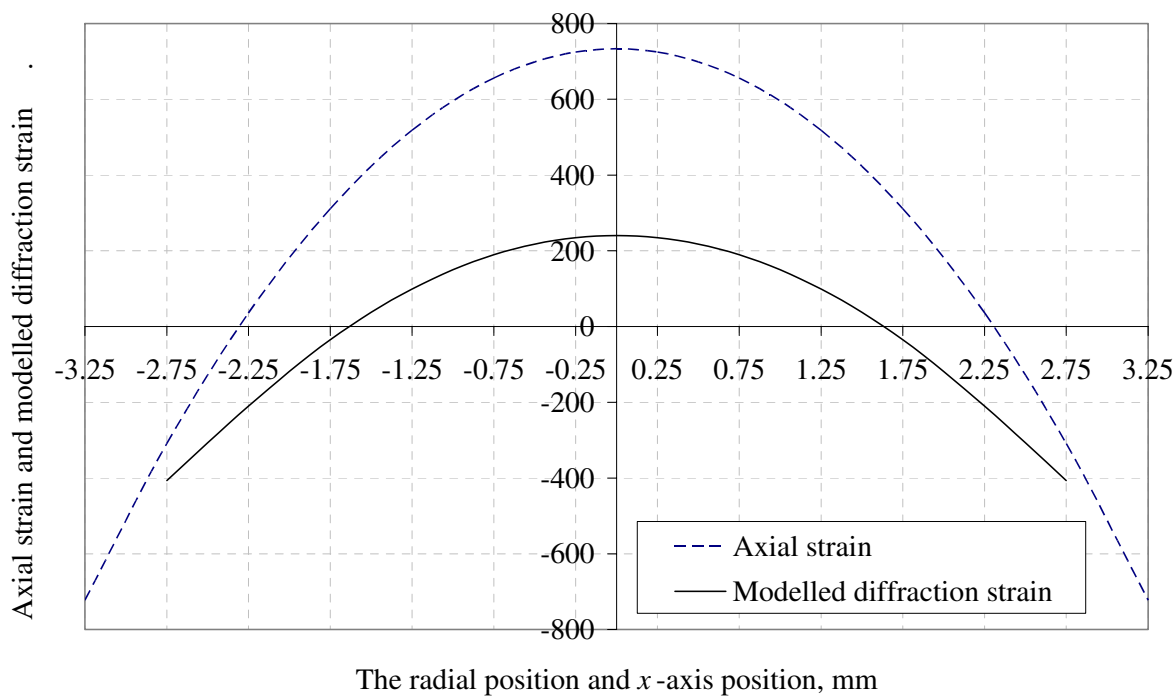} }
\caption{Comparison between the reconstructed axial strain within the sample (dashed line) and the modelled diffraction strain (continuous solid curve), as a function of the radial position, $r$, and the position within the $x$-scan centred on the sample axis of symmetry.
}
\label{fig:six}
\end{figure}

Fig.\ref{fig:six} shows the comparison between the reconstructed axial strain within the sample, plotted as the function of the radial distance from the cylindrical axis, and the modelled diffraction strain, that is plotted as a function of the position within the $x$-scan across the sample that is also centred on the cylindrical axis. The distinction between the two curves is immediately clear. 

Firstly, it is obvious that the axial strain is greater in magnitude than the measured diffraction strain. This happens because diffraction strain is in fact an average of the strain distribution within the sampling volume, and thus necessarily has a smaller range of variation and 'misses' the peak strain values. In the plot the axial strain curve is intentionally extended to the surface of the specimen, $r$ = 3.25mm, whereas the diffraction strain curve is terminated at $x$ = 2.75mm, i.e. the last point for which the gauge volume is still fully immersed in the sample.

The most tensile and most compressive axial strain values are both close in magnitude to about 750 microstrain. By way of estimate, assuming Young's modulus of about 200 GPa, the tensile stress at the axis of the cylindrical sample may therefore be estimated to be close to 150 MPa, and a similar magnitude compressive stress is acting at the surface. These stresses possess sufficient magnitude in order to affect the fatigue life of a component, and must be taken into account. Diffraction strain tomography provides a useful tool for their determination, and can be developed further to apply to more complex internal strain distributions.

\section*{Acknowledgements}

The authors would like to thank Dr Simon Jacques, Professor Paul Barnes and Professor Bob Cernik for a discussion that provoked the development of ideas presented in this note.

% Run this to generate the bibliography file, then switch off

%\bibliographystyle{plain}
%\bibliography{stom}% Produces the bibliography via BibTeX.

\end{document}